\documentclass[twoside,twocolumn,12pt]{article}
\usepackage{extsizes}
\usepackage[super,sort&compress,comma]{natbib} 
\usepackage[version=4]{mhchem}
\usepackage{amssymb}
\usepackage[left=1.5cm, right=1.5cm, top=1.785cm, bottom=2.0cm]{geometry}
\usepackage{balance}
\usepackage{times,mathptmx}
\usepackage{sectsty}
\usepackage{graphicx} 
\usepackage{lastpage}
\usepackage[format=plain,justification=justified,singlelinecheck=false,font={stretch=1.125,small,sf},labelfont=bf,labelsep=space]{caption}
\usepackage{float}
\usepackage{fancyhdr}
\usepackage{fnpos}
\usepackage[english]{babel}
\usepackage{array}
\usepackage{droidsans}
\usepackage{charter}
\usepackage[T1]{fontenc}
\usepackage[usenames,dvipsnames]{xcolor}
\usepackage{setspace}
\usepackage[compact]{titlesec}
\usepackage{url}
\usepackage[mathscr]{euscript}
\usepackage[pagewise]{lineno}
\usepackage{caption}
\usepackage{setspace}
\usepackage[normalem]{ulem}
\usepackage{amsmath}
\usepackage{chemformula}

\definecolor{cream}{RGB}{222,217,201}

\begin{document}

\pagestyle{fancy}
\thispagestyle{plain}
\fancypagestyle{plain}{\renewcommand{\headrulewidth}{0pt}}

\makeFNbottom
\makeatletter
\renewcommand\LARGE{\@setfontsize\LARGE{15pt}{17}}
\renewcommand\Large{\@setfontsize\Large{12pt}{14}}
\renewcommand\large{\@setfontsize\large{10pt}{12}}
\renewcommand\footnotesize{\@setfontsize\footnotesize{7pt}{10}}
\makeatother

\renewcommand{\thefootnote}{\fnsymbol{footnote}}
\renewcommand\footnoterule{\vspace*{1pt}%
\color{cream}\hrule width 3.5in height 0.4pt \color{black}\vspace*{5pt}} 
\setcounter{secnumdepth}{5}

\makeatletter 
\renewcommand\@biblabel[1]{#1}            
\renewcommand\@makefntext[1]%
{\noindent\makebox[0pt][r]{\@thefnmark\,}#1}
\makeatother 
\renewcommand{\figurename}{\small{Fig.}$\sim$}
\sectionfont{\sffamily\Large}
\subsectionfont{\normalsize}
\subsubsectionfont{\bf}
\setstretch{1.125}
\setlength{\skip\footins}{0.8cm}
\setlength{\footnotesep}{0.25cm}
\setlength{\jot}{10pt}
\titlespacing*{\section}{0pt}{4pt}{4pt}
\titlespacing*{\subsection}{0pt}{15pt}{1pt}

\fancyfoot{}
\fancyfoot[RO]{\footnotesize{\sffamily{\thepage}}}
\fancyfoot[LE]{\footnotesize{\sffamily{\thepage}}}
\fancyhead{}
\renewcommand{\headrulewidth}{0pt} 
\renewcommand{\footrulewidth}{0pt}
\setlength{\arrayrulewidth}{1pt}
\setlength{\columnsep}{6.5mm}
\setlength\bibsep{1pt}

\makeatletter 
\newlength{\figrulesep} 
\setlength{\figrulesep}{0.5\textfloatsep} 

\newcommand{\topfigrule}{\vspace*{-1pt}%
\noindent{\color{cream}\rule[-\figrulesep]{\columnwidth}{1.5pt}} }

\newcommand{\botfigrule}{\vspace*{-2pt}%
\noindent{\color{cream}\rule[\figrulesep]{\columnwidth}{1.5pt}} }

\newcommand{\dblfigrule}{\vspace*{-1pt}%
\noindent{\color{cream}\rule[-\figrulesep]{\textwidth}{1.5pt}} }

\makeatother

\twocolumn[
\begin{@twocolumnfalse}
	\vspace{3cm}
	\sffamily
	\noindent\LARGE{\textbf{The weakest link bridging germinal center B cells and follicular dendritic cells limits antibody affinity maturation\\}} \\
	
	\noindent\Large{Rajat Desikan$^{\ast}$\textit{$^{a}$}$^{\dag}$, Rustom Antia\textit{$^{b}$} and Narendra M. Dixit$^{\ast}$\textit{$^{a,c}$}} \\ \\ \\ \\ \\ \\ \\ \\
	
\end{@twocolumnfalse} \vspace{0.6cm}

]

\renewcommand*\rmdefault{bch}\normalfont\upshape
\rmfamily
\section*{}
\vspace{-1cm}

\footnotetext{\textit{$^{a}$Department of Chemical Engineering, $^{b}$Department of Biology, Emory University, Atlanta, Georgia, United States of America, $^{c}$Centre for Biosystems Science and Engineering, Indian Institute of Science, Bengaluru, India, 560012}}
\footnotetext{$^{\dag}$Present address: Akamara Biomedicine Pvt. Ltd., No. 465, Patparganj Industrial Area, Delhi 110092, India}
\footnotetext{$^{\ast}$Corresponding authors, E-mail: rajatdesikan@gmail.com; narendra@iisc.ac.in}

\onecolumn
\renewcommand{\figurename}{Figure}

\clearpage
\newpage
\doublespacing
\raggedbottom

\section*{Abstract}

The affinity of antibodies (Abs) produced \textit{in vivo} for their target antigens (Ags) is typically well below the maximum affinity possible. Nearly 25 years ago, Foote and Eisen explained how an `affinity ceiling' could arise from constraints associated with the acquisition of soluble antigen by B cells. However, recent studies have shown that B cells in germinal centers (where Ab affinity maturation occurs) acquire Ag not in soluble form but presented as receptor-bound immune complexes on follicular dendritic cells (FDCs). How the affinity ceiling arises in such a scenario is unclear. Here, we argue that the ceiling arises from the weakest link of the chain of protein complexes that bridges B cells and FDCs and is broken during Ag acquisition. This hypothesis explains the affinity ceiling realized \textit{in vivo} and suggests that strengthening the weakest link could raise the ceiling and improve Ab responses.
    
\newpage

\section*{Main}
Antibodies (Abs) produced \textit{in vivo} later in an infection tend to have higher affinities for their target antigen (Ag) than those produced earlier on.\cite{Eisen1964, Nussenzweig2012} This phenomenon, termed antibody affinity maturation\cite{Nussenzweig2012, Cyster2019}, has long been recognized as a hallmark of humoral immunity and demonstrates its ability to `learn' to recognize its target better with time. When challenged with simple Ags such as haptens conjugated to proteins, mice produced Ag-specific Abs with affinities, quantified using the equilibrium association constant ($K_a$), that increased from $10^5-10^6$ M\textsuperscript{-1} a week after challenge to $10^7-10^8$ M\textsuperscript{-1} in a few months.\cite{Eisen1964} A similar rise is seen for complex Ags: The affinity of broadly neutralizing Abs (bNAbs), which target conserved regions on the HIV-1 envelope, rose gradually from $10^4-10^6$ M\textsuperscript{-1} to $10^8-10^9$ M\textsuperscript{-1} in a few years in HIV-1 infected individuals (Figure 1\textit{a}).\cite{Liao2013, Bonsignori2016} The mean affinity of Abs against influenza hemagglutinin were similarly found to rise from $\sim10^4$ M\textsuperscript{-1} and exceed $10^6$ M\textsuperscript{-1}.\cite{Kuraoka2016} Ab affinities can thus rise many orders of magnitude during the course of an infection.  

This rise, however, is not unabated. The affinity eventually saturates. The maximum affinities realized \textit{in vivo} correspond to $K_a \sim10^{10}-10^{12}$ M\textsuperscript{-1} (Refs.{\cite{Foote1995,Foote2000,Chan2009}}) (Figure 1\textit{b}). Intriguingly, this saturating affinity is far below the maximum affinity realizable between proteins or Ab-Ag pairs. The maximum $K_a$ recorded between proteins, that for the avidin-biotin interaction, is $\sim 10^{15}$ M\textsuperscript{-1} (Ref.{\cite{Green1975}}). Abs with $K_a > 10^{13}$ M\textsuperscript{-1} for their target Ags have been realized \textit{in vitro} using directed evolution.\cite{Boder2000} Affinity maturation \textit{in vivo} thus appears to hit an `affinity ceiling'.\cite{Foote2000} The potency with which Abs neutralize their targets is positively correlated with their affinities for the targets.{\cite{Maynard2002,Carter2006,Brandenberg2017}} The ceiling thus potentially limits the potency of the endogenous Ab response. Unraveling the origins of the ceiling and devising ways to manipulate it would have implications not only for our understanding of the Ab response but also for optimizing vaccinations and immunotherapies for infections and cancer {\cite{Lu2018, Carter2018}}.

In an insightful commentary nearly 25 years ago, Foote and Eisen presented an explanation of the affinity ceiling considering the scenario wherein B cells acquired `soluble' Ag.\cite{Foote1995} As we describe in more detail later, they argued that limits to the association and dissociation rate constants, $k_{\rm on}$ and $k_{\rm off}$, of the Ab-Ag interactions, were set by molecular diffusion and receptor endocytosis, respectively, which could not be breached by mutations in the Ag binding regions of the Abs, thus giving rise to the affinity ceiling. In particular, they suggested that the rate of encounter by diffusion set an upper bound on $k_{\rm on}$, and that if $k_{\rm off}$ was low enough that the Ag would get internalized prior to its dissociation from the B cell, there would be little selection for further reductions in $k_{\rm off}$. \newline

\begin{figure}[H]
	\centering
	\includegraphics[width=\textwidth]{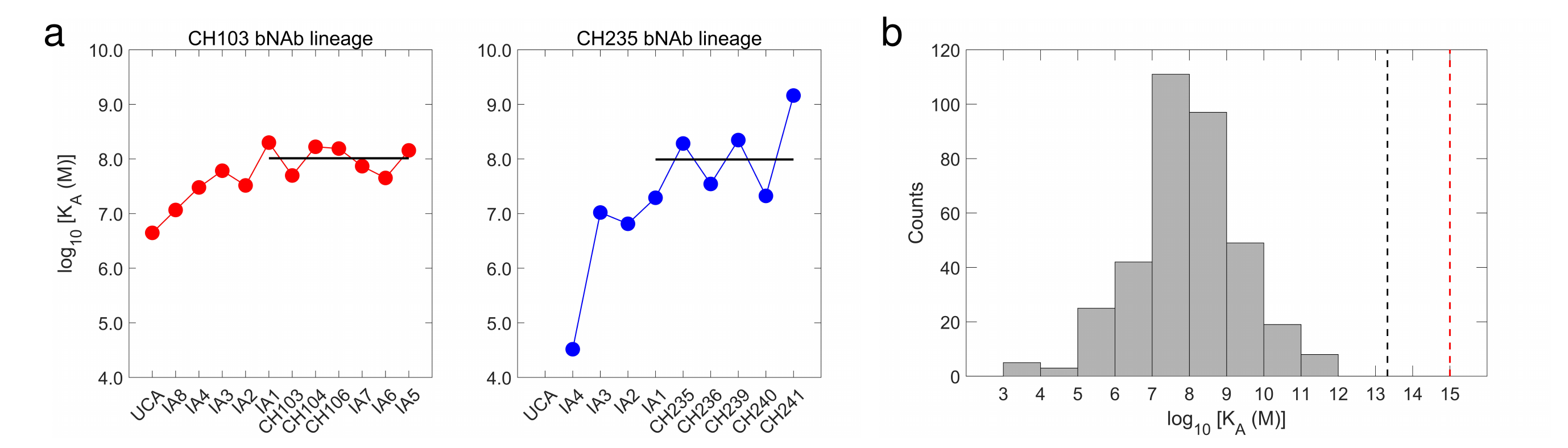}
	\caption{\textbf{The antibody affinity ceiling.} (\textbf{a}) Affinity maturation of two anti-HIV-1 broadly neutralizing antibody lineages, CH103 (red) and CH235 (blue) {\cite{Bonsignori2016}}. Horizontal black line indicates the mean $K_a$ of the last six members of both lineages. (\textbf{b}) Histogram of antibody-antigen affinities reported in the structural antibody database (SAbDAb){\cite{Dunbar2014}}. Vertical black and red dashed lines correspond to $K_a$ of $2.1 \times 10^{13}$ M\textsuperscript{-1} and $10^{15}$ M\textsuperscript{-1}, respectively, representing the highest Ab-Ag affinity \textit{in vitro}\cite{Boder2000} and the highest protein-protein affinity\cite{Green1975} reported, illustrating the gap between the affinities realizeable and those realized \textit{in vivo}, limited by the affinity ceiling.}
\end{figure}

In recent years, it has become clear that affinity maturation occurs in specialized regions, namely, germinal centers (GCs), in lymph nodes and secondary lymphoid organs.  Advances in experimental techniques, especially intravital imaging technologies that allow direct visualization of B cells within GCs, have established that B cells acquire Ag not in soluble form but presented as Ab-Ag immune complexes (ICs) on the surfaces of follicular dendritic cells (FDCs) in GCs\cite{Nussenzweig2012,Kwak2019,Cyster2019}, challenging the arguments of Foote and Eisen. An explanation of the affinity ceiling that is consistent with the modern view of the GC reaction is warranted. In this Perspective, we suggest that the ceiling arises from limits to the strength of a chain of protein complexes formed between GC B cells and FDCs for Ag acquisition. The limit is set by the weakest link in this chain.\newline 
   
\section*{Affinity maturation}
Affinity maturation occurs in GCs, which are temporary structures that get assembled in secondary lymphoid organs following an infection (Figure 2).\cite{Nussenzweig2012} Here B cells, termed GC B cells, continuously evolve and get selected based on the affinity of their B cell receptors (BCRs) for their target Ag. We briefly describe the GC reaction\cite{Nussenzweig2012, Cyster2019, Kwak2019}. GCs are divided anatomically into two regions, a light zone and a dark zone (Figure 2). GC B cells are pro-apoptotic by default and must receive two signals in the light zone to survive.\cite{Mayer2017} First, they must acquire Ag. Second, they must present the acquired Ag as peptides on major histocompatibility complex II (pMHCII) to T follicular helper ($T_{fh}$) cells and receive help from them. When these two signals are received, the B cells are rescued and a majority migrate to the dark zone, where they proliferate and experience mutations in their BCR, or Ab, genes. (Abs are secreted forms of BCRs; the two have identical Ag binding regions or `paratopes'.) Following proliferation and mutation in the dark zone B cells migrate back to the light zone where they are subjected to the same survival pressures again. The likelihood of the receipt of survival signals depends on the affinity of the BCR for Ag. Each B cell expresses a single kind of BCR. The greater the affinity, the greater is the chance of Ag acquisition and hence also of receiving $T_{fh}$ cell help. Consequently, progressively, B cells with increasing affinity for the target Ag are selected, resulting in affinity maturation. A small percentage of selected GC B cells continuously differentiate into plasma cells (and memory B cells), exit the GC and produce Abs (Figure 2), explaining the observed increase in the affinity of the Abs with time.\newline

\section*{The soluble Ag scenario (Foote and Eisen model)}
Why does affinity maturation saturate? Foote and Eisen answered this question for the scenario wherein B cells acquire soluble Ag.\cite{Foote1995} They argued that the affinity ceiling would arise from independent limits on the association and dissociation rate constants,  $k_{\rm on}$ and $k_{\rm off}$, respectively, of the Ab-Ag interactions.\cite{Foote1995} They reasoned that $k_{\rm on}$ is limited by the diffusion of Abs to the soluble Ag, as has been verified experimentally\cite{Raman1992, Northrup1992}, and could be set to a maximum of $\approx 10^5-10^{6}$ M\textsuperscript{-1} s\textsuperscript{-1} (Refs.{\cite{Foote1995,Raman1992,Northrup1992}}). $k_{\rm on}$ is thus not sensitive to the specific inter-molecular interactions, defined by the conformations, electrostatics, hydrophobicity, and other features of the proteins. The difference in the affinity between different Abs for a given Ag must thus arise from different values of $k_{\rm off}$. This difference is expected to be manifested in B cell selection. The smaller the $k_{\rm off}$, the more stable would be the BCR-Ag complex on the B cell surface, which could provide an advantage to the B cell in terms of a longer stimulation of BCR signaling and/or a higher probability of Ag uptake. Foote and Eisen recognized that this selective advantage would cease once $k_{\rm off}$ dropped to a value close to the rate of internalization of BCR-Ag complexes by B cells. All BCR-Ag complexes lasting longer (\textit{i.e.,} with smaller values of $k_{\rm off}$) would be internalized and processed similarly, leaving little room for affinity discrimination. Thus, the internalization, or endocytosis, rate, $\sim 10^{-3}-10^{-4}$ s\textsuperscript{-1} (Ref. \cite{Watts1988}), set a lower bound on $k_{\rm off}$. Combining the two limits --- the upper bound on $k_{\rm on}$ and the lower bound on $k_{\rm off}$ --- yielded an estimate of the affinity ceiling: $k_{\rm on}/k_{\rm off}\approx10^{10}$ M\textsuperscript{-1}.

This estimate is surprisingly close to the ceiling observed.\cite{Roost1995, Foote1995}  Further, Batista and Neuberger\cite{Batista1998} experimentally validated the above rationale for the ceiling. They exposed B cells to soluble Ag \textit{in vitro} and assessed their ability to stimulate Ag-specific CD4 T cells in co-culture by measuring the amount of interleukin 2 (IL2) secreted by the T cells. Remarkably, the stimulation was sensitive to $k_{\rm off}$ when the BCR-Ag bound lifetime was $<12$ min, close to the internalization timescale\cite{Watts1988}, but became independent of $k_{\rm off}$ once the lifetime exceeded $12$ min. 

The problem of the affinity ceiling thus appeared settled. Recent experimental advances, however, have resurrected the problem. Our view of the GC today differs fundamentally from the scenario on which the arguments of Foote and Eisen are predicated.\cite{Cyster2019, Kwak2019} Given that GC B cells do not acquire soluble Ag in GCs, the limits on $k_{\rm on}$ and $k_{\rm off}$ that define the affinity ceiling \textit{in vivo} need to be re-evaluated. What then defines the ceiling?\newline

\begin{figure}[!htp]
	\centering
	\includegraphics[width=0.9\textwidth]{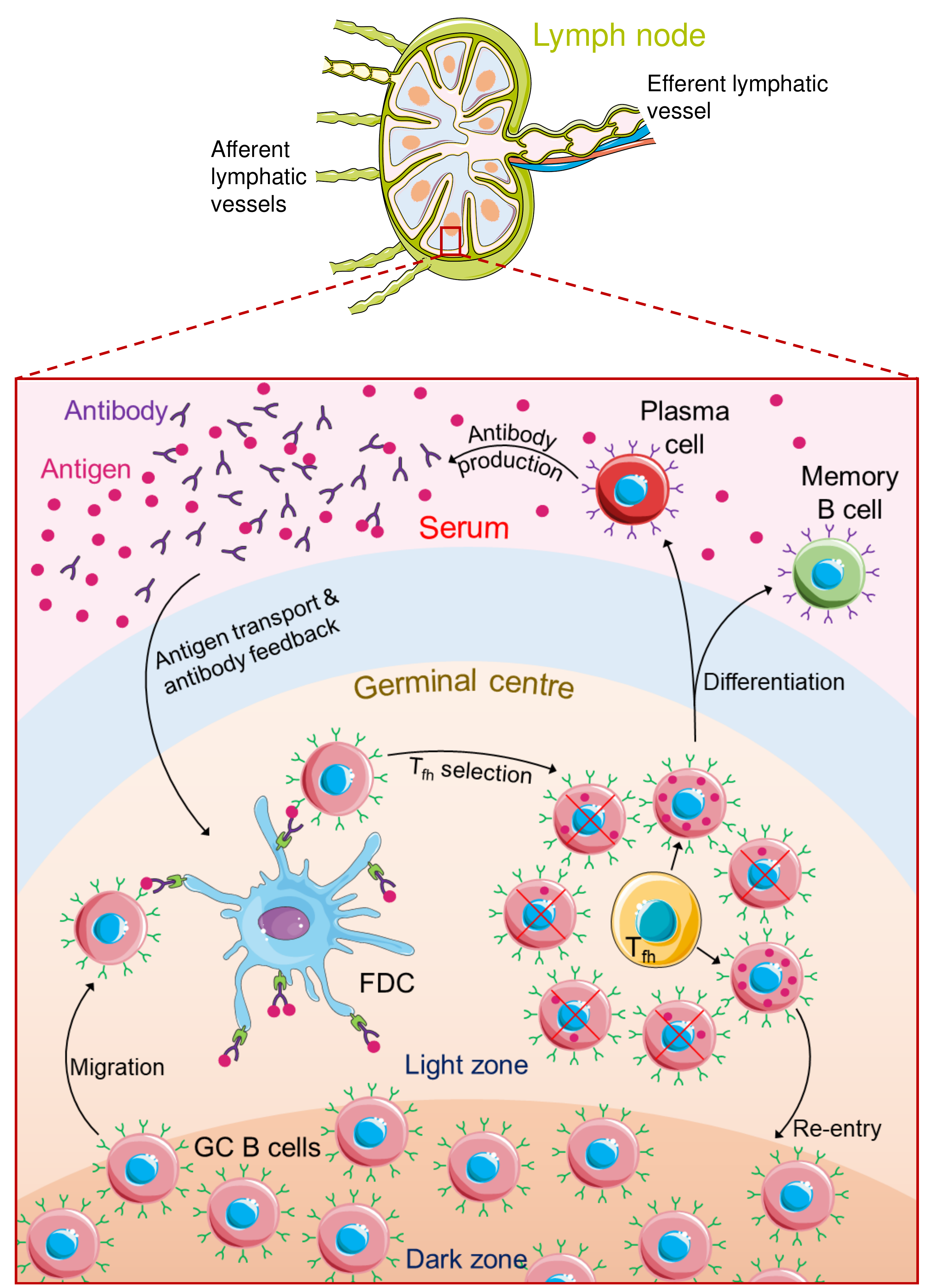}
	\caption{(Continued on the following page.)}
\end{figure}

\addtocounter{figure}{-1}
\begin{figure}[!htp]
	\centering
	\caption{\textbf{B cell evolution and affinity maturation in GCs.} Cartoon of a lymph node with afferent and efferent lymphatic vessels that facilitate Ag/Ab transport. GCs (zoomed below) are specialized, temporarily assembled B cell rich structures, anatomically partitioned into dark and light zones. GC B cells in the light zone interact with Ag-presenting FDCs and acquire Ag with probabilities increasing with the affinity of their BCRs for the Ag. They then receive T$_{fh}$ cell help with a probability proportional to the amount of Ag acquired and survive. A majority of the GC B cells thus selected enter the dark zone, where they proliferate and experience mutations of their Ab genes. They then migrate back to the light zone and are subjected to the same selection process, thus resulting in gradual affinity maturation. A small percentage of the selected GC B cells differentiates into plasma cells or memory B cells and exits the GC. Abs produced by plasma cells in the serum may feedback into the GC, render Ag acquisition more difficult, and increase the B cell selection stringency.}
\end{figure}

\section*{Alternative explanations}

We examined several alternative explanations. 

\subsection*{\textit{Inadequate time}} Infections can get cleared rapidly and offer inadequate time for B cells to evolve and Ab affinities to rise much beyond the ceiling. Note that the ceiling refers to the highest affinities seen \textit{in vivo}. In many situations, the maximum affinities realized are far below the ceiling (Figure 1\textit{b}). Shortage of time, however, appears not to be a satisfactory explanation for two reasons. First, the ceiling is seen also with chronic infections such as HIV-1, which can last tens of years. In fact, the evolution of HIV-1 bNAbs, which target conserved regions on the virus, is known to take years.\cite{Liao2013} Still, the affinities of bNAbs do not breach the ceiling; rather, they tend to be much lower than the ceiling (for example, see Figure 1\textit{a}). Second, on rare occasions, Abs with affinities close to the ceiling can arise, due to random mutations in the Ab genes, soon after the onset of infection. Yet, as the infection progresses, the affinities do not rise above the ceiling.\cite{Roost1995}  

\subsection*{\textit{Antigen evolution}} A persistent evolutionary arms race between pathogens or tumor cells and the host immune system is characteristic of infections with rapidly mutating pathogens such as HIV and hepatitis C virus\cite{Nourmohammad2016} and cancer\cite{Spranger2018}. It is conceivable in such a scenario that Ab evolution, due to B cell selection, increases Ab affinity for a target Ag, whereas Ag evolution, via mutation, compromises the affinity by altering the target. A balance between these competing effects may define the ceiling. This explanation, however, fails to describe how the ceiling arises with simple, non-mutating Ag such as haptens, nor does it explain why \textit{in vitro} selection can allow antibody affinity to breach this ceiling\cite{Boder2000}.  
	
\subsection*{\textit{Insufficient GC B cells}} In chronic infections, a larger fraction of selected GC B cells has been argued recently to be terminally differentiated into plasma and memory cells, which may leave insufficient numbers of GC B cells to efficiently mutate and continue the GC reaction.\cite{Staupe2019} While this phenomenon may contribute to the ceiling in chronic infection settings, it does not explain the ceiling seen soon after immunization \sout{in some settings}\cite{Roost1995}. 

\subsection*{\textit{Structural limitations}} One could argue that intrinsic limitations in Ab structures may preclude the realization of higher affinities for Ag than the ceiling. This possibility has been negated, as mentioned above, by \textit{in vitro} studies that have generated Abs with affinities much higher than the ceiling\cite{Boder2000}.

Based on these considerations, it followed that the origins of the ceiling were likely to be in processes more directly associated with Ag acquisition and GC B cell selection leading to affinity maturation \textit{in vivo}. We focussed on these processes next. Below, we summarize recent advances in our understanding of these processes based on which we constructed the weakest link hypothesis.\newline      

\section*{Recent advances in our understanding of affinity maturation}

\subsection*{\textit{Ag acquisition}} GC B cells sample Ag presented as ICs attached non-covalently to Fc or complement receptors on antigen presenting cells (APCs), particularly FDCs, in GCs.\cite{Batista2001,Fleire2006} Soluble Ag is hardly sampled. B cells form contacts with the FDCs through the BCR-IC interactions (Figures 2 and 3) as well as other adhesion receptors such as integrins.\cite{Batista2001,Kwak2019} The B cell-APC contact region, termed the `immunological synapse', is enriched in BCRs and other proteins.\cite{Batista2001,Kwak2019} B cells exert a mechanical pulling force on the BCR to acquire Ag, which renders Ag affinity discrimination by B cells a function of the mechanical properties of the underlying substrate\cite{Natkanski2013,Spillane2017}: When the substrate is highly flexible, Ag acquisition is readily accomplished with the force, largely independently of BCR-Ag affinity, rendering affinity discrimination poor. When the membrane is moderately flexible, the force ruptures the weak BCR-Ag complexes, but allows sustained bond formation and Ag acquisition with strong BCR-Ag complexes, facilitating affinity discrimination. Finally, when the substrate is extremely rigid, B cells switch to a biochemical pathway, hydrolyzing the Ag using enzymes secreted into the synaptic region and then acquiring the Ag.\cite{Yuseff2011, Spillane2017} FDCs, which form the predominant substrate for Ag acquisition in GCs, appear to be moderately flexible.\cite{Spillane2017}

\subsection*{\textit{Affinity discrimination}} GC B cells are significantly different from na{\"i}ve B cells in the way they acquire Ag from APCs, which renders them more adept at affinity discrimination\cite{Kwak2018}: They express fewer BCRs, as well as fewer integrins and other anchoring proteins involved in synapse formation between B cells and APCs. Unlike na{\"i}ve B cells, which spread over a surface presenting Ag and form uniform contacts, GC B cells form highly dynamic, punctate contacts, which involve long, pod-like extensions of their membranes, enriched in F-actin and ezrin molecules, which are responsible for membrane shape changes. GC B cells exert a force on the BCR-Ag complexes through the BCRs in these puncta that is much larger than the force exerted by na{\"i}ve B cells. When placed in contact with surfaces displaying low affinity Ag, they form short-lived contacts, whereas the contacts quickly stabilize when the surface presents high affinity Ag. GC B cells also aquire a lot more high affinity Ag than low affinity Ag. The force applied, thus, appears to break the low affinity BCR-Ag complexes more readily, preventing their internalization. 

\subsection*{\textit{Ag internalization}} The process of Ag internalization by GC B cells appears to be distinct from na{\"i}ve B cells\cite{Kwak2018}: na{\"i}ve B cells internalize Ag into endosomes near the immunological synapse, whereas GC B cells tend to traffic Ag on the cell membrane to sites distant from the synapse before internalization. Further, GC B cells express far fewer of the proteins SNX9 and SNX18, involved in endocytosis, than na{\"i}ve B cells. Besides, while the latter proteins are concentrated in the synapse in na{\"i}ve B cells, such concentration is not evident with GC B cells.  

\subsection*{\textit{Ab feedback}} The selection stringency for B cells in the GCs increases steadily due to Ab feedback\cite{Zhang2013, Zhang2016, Garg2019}: As the GC reaction proceeds, Abs produced by recently differentiated plasma cells can traffic back to the GC. They can bind Ag presented on FDCs and mask them from B cells, rendering Ag acquisition difficult. Further, if their affinity for Ag is higher than the Abs presenting Ag as ICs on FDCs, they can replace the latter Abs and themselves present Ag as ICs. BCRs must form ternary complexes with ICs on FDCs and then extract Ag from the ICs, which would require the dissociation of the IC\cite{Batista2000}. As the affinity of the feedback Abs rises due to affinity maturation, Ag extraction by BCRs becomes increasingly difficult. Indeed, passive immunization with Abs of high affinity for Ag led to an increased frequency of GC collapse\cite{Zhang2013}, indicative of an insufficient population of selected GC B cells. 

Based on the above advances, we reasoned that the ceiling was likely to arise from limitations to affinity discrimination at the Ag acquisition stage following the B cell-FDC contact. The weakest link hypothesis emerged from this reasoning.\newline
   
\section*{The weakest link hypothesis}

We focus on a BCR in contact with its target Ag in the GC B cell-FDC contact region (Figure 3\textit{a}). The Ag is in an IC presented attached to an Fc or CR2 receptor on the FDC surface. The BCR is subjected to a force by the contractile motion of the B cell surface.\cite{Spillane2017, Kwak2018} This force is transmitted across the chain of protein complexes that links the B cell to the FDC. This chain involves the following complexes: BCR-Ag, Ag-Ab (IC), and Ab-Fc$\gamma$RIIB receptor (or Ab-CR2 receptor). In addition, the BCR is anchored to the B cell membrane and the Fc$\gamma$RIIB (or CR2) receptor to the FDC membrane. Because internalization of Ag happens typically after the Ag is trafficked to sites distant from the synapse along the B cell membrane\cite{Kwak2018}, the chain is expected to snap due to the force to allow such trafficking. We expect the chain to snap at its weakest link. 

\begin{figure}[!htp]
	\centering
	\includegraphics[width=0.9\textwidth]{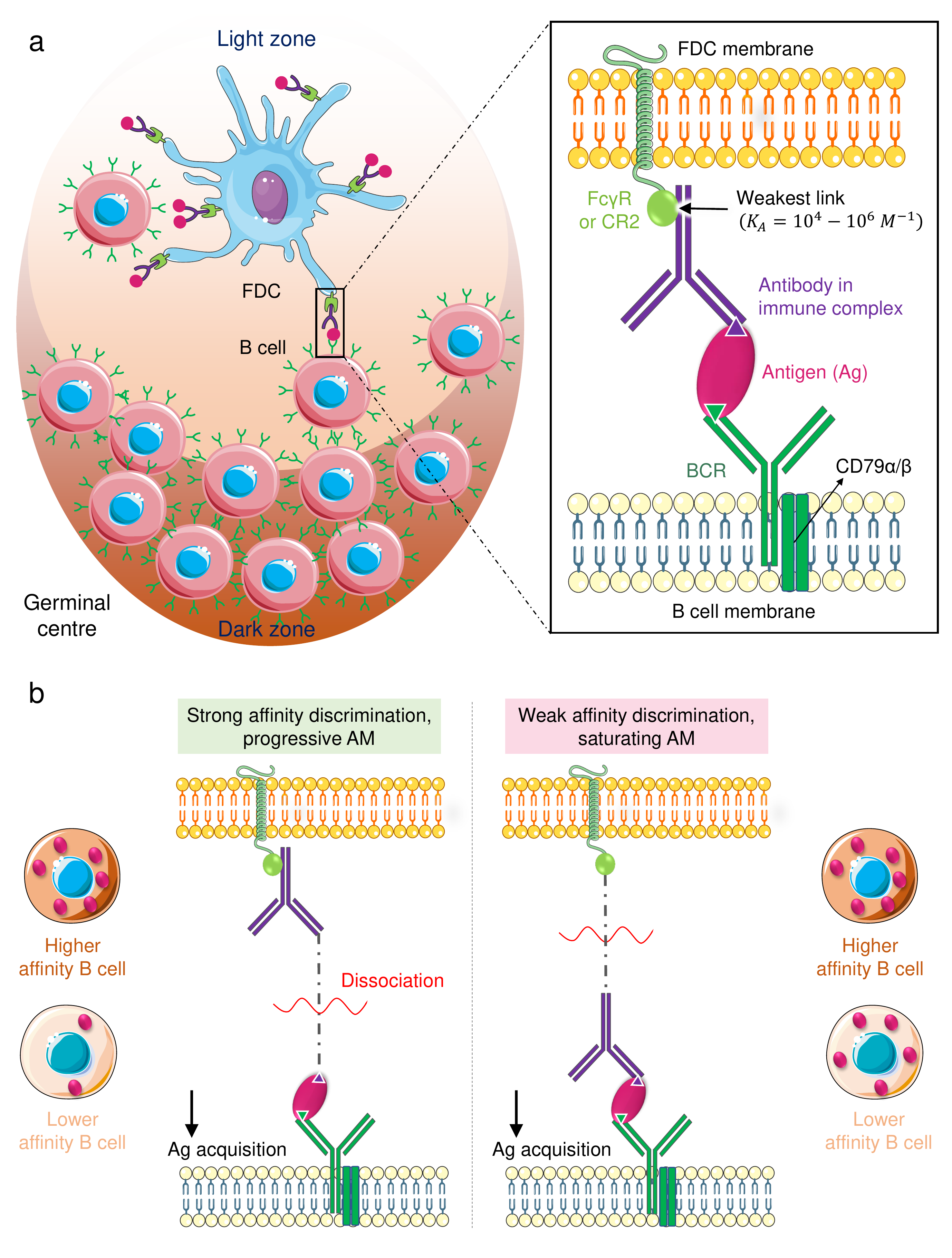}
	\caption{(Continued on the following page.)}
\end{figure}

\addtocounter{figure}{-1}
\begin{figure}[!htp]
	\centering
	\caption{\textbf{The B cell-FDC tug-of-war, weakest link in the chain, and affinity maturation regimes.} (\textbf{a}) GC B cells interact with Ag-presenting FDCs in the light zone of the GC. The B cell-FDC contact (right zoom) is through a chain of non-covalently bound protein-protein and proteo-lipid complexes: FDC membrane with Fc$\gamma$RIIB/CR2 receptors; Fc$\gamma$RIIB/CR2 receptors with the presenting Ab in the IC; the presenting Ab with Ag; Ag with BCR; and BCR with the B cell membrane. The Ab/BCR interactions with the Ag occur at the respective epitopes (triangles). Energetically, the weakest link in this interacting chain upon sufficient Ab affinity maturation is the Fc$\gamma$RIIB/CR2--Ab complex. (\textbf{b}) (Left) Progressive affinity maturation. Ab-Ag interaction is the weakest link and Ag acquisition by B cells happens by BCRs breaking the Ab-Ag complex. B cells with higher affinity BCRs will thus acquire Ag more efficiently and typically end up with higher amounts of Ag than B cells with lower affinity BCRs. (Right) Saturating affinity maturation. Fc$ \gamma $RIIB/CR2--Ab interaction is the weakest link, and B cells acquire Ag by breaking the Fc$ \gamma $RIIB/CR2--Ab complex. The amount of Ag acquired thus ceases to depend on the affinity of BCRs for Ag, precluding further affinity maturation.}
\end{figure}

To determine the weakest link, we examine the binding free energies, \textit{i.e., }$ \Delta G $'s, of all the links in the chain. The lower (more negative) is the value of $\Delta G$, the stronger is the link. We assume that the BCR is tightly anchored into the B cell membrane and is unlikely to be the weakest link. If the BCR were to dislodge from the B cell membrane, no Ag acquisition would occur and the B cell would not survive. The $ \Delta G $'s of anchoring Fc or CR2 receptors in the FDC membrane have not been measured. Theoretical calculations of `water-to-membrane' free energies are reported in the membranome database{\cite{Lomize2017,Lomize2018}}, which provide estimates. Accordingly, the $ \Delta G $ of Fc receptor anchoring in the FDC membrane is -106 kJ/mol and that of the CR2 receptor is -182 kJ/mol. The affinities of the Fc$ \gamma $RIIB receptor with the Fc region of Abs has been measured for all the four IgG sub-classes.{\cite{Bruhns2009}} The affinities lie in the range $K_A=0.25-2.5\times10^{5}$ M\textsuperscript{-1}, which would correspond to $ \Delta G $'s in the range -26 kJ/mol to -32 kJ/mol at 37$^0$C. Similarly, the binding of CR2 to its ligands such as C3d is well understood structurally, and surface plasmon resonance experiments have indicated that a monovalent reaction model (1:1 stoichiometry of CR2:C3d) best fits the binding kinetics of CR2 with C3d.{\cite{vandenElsen2011}} The reported affinity is $\sim 2\times10^6$ M\textsuperscript{-1}, which corresponds to a $ \Delta G $ of -37 kJ/mol. Note that multiple C3d molecules can bind to a single IC and interact with complement receptors, potentially increasing the overall affinity. We look finally at the $ \Delta G $ of the Ab-Ag interaction. At the start of the GC reaction, the affinities can be low, $\sim 10^{5}$ M\textsuperscript{-1}, yielding $ \Delta G $ of -29 kJ/mol. As the reaction proceeds, the affinity can rise up to $\sim 10^{7}-10^{12}$ M\textsuperscript{-1}, corresponding to $ \Delta G $ of -42 kJ/mol to -71 kJ/mol.

It follows from the  $ \Delta G $'s that at the start of the GC reaction, the weakest link could be the Ab-Ag interaction. Thus, when a B cell with a BCR of a higher affinity for Ag than the Ab in the IC presenting the Ag encounters the FDC, the chain would snap at the Ab-Ag interaction, dissociating the IC and resulting in Ag acquisition by the B cell (progressive AM regime in Figure 3\textit{b}). As the GC reaction progresses, Ab feedback results in the Ab-Ag interaction becoming stronger relative to the Fc$ \gamma $RIIB/CR2--Ab interaction. When the strength of Ab-Ag interaction exceeds that of the Fc$ \gamma $RIIB/CR2--Ab interaction, the latter link becomes the weakest and the chain snaps there (saturating AM regime in Figure 3\textit{b}). 

The Fc$ \gamma $RIIB/CR2--Ab interaction is removed from the Ag and is thus not likely, assuming weak allosteric effects, to be sensitive to Ab mutations that can increase Ag-Ab affinities.  Thus, all BCRs with affinities higher than the Fc$ \gamma $RIIB/CR2--Ab interaction will acquire Ag by snapping the Fc$ \gamma $RIIB/CR2--Ab complex, leaving no selective advantage for BCRs with higher affinities. The Ag-Ab affinity ceiling is thus determined by the affinity of the Fc$ \gamma $RIIB/CR2--Ab complex.\newline
    
\section*{Quantifying the affinity ceiling}

We now consider different potential scenarios, or configurations, of Ag presentation and acquisition, defined by the number of tethers holding the Ag attached to the FDC and the number of BCRs tugging it away, to quantify the ceiling.  

\subsection*{Bivalent antigen}
We consider first the simplest scenario where affinity maturation occurs for an Ag that has two identical Ab binding sites that are accessible simultaneously but are sufficiently separated that each site can be accessed by a single Abs/BCRs. We assume further that B cells of a single lineage evolve and result in affinity maturation. This scenario, illustrated in Figure 4\textit{a}, is similar to Figure 3 but with the purple and green epitopes being identical as are the lineages of the corresponding Abs/BCRs. When the ceiling is reached, the presenting Ab in the IC as well as the BCR on the interacting B cell will have affinities for the Ag just above the strength of the weakest link, the Fc$ \gamma $RIIB-Ab or the CR2-Ab complex, $\kappa$, the latter in the range of $10^{4}-10^6$ M\textsuperscript{-1}. The ceiling would thus be $10^{4}-10^6$ M\textsuperscript{-1}, corresponding to the snapping of a single Fc$\gamma$RIIB-Ab or CR2-Ab complex (Figure 3\textit{b}).

We next consider the scenario where the Ag is bivalent but the epitopes are distinct, i.e., the purple and green epitopes in Figure 4\textit{b} are not identical. Affinity maturation would now have to occur simultaneously for the two epitopes, with two distinct lineages of B cells - expressing purple and green BCRs, respectively - evolving for the two epitopes. This is because a B cell with a green BCR lineage, targeting the green epitope, can acquire Ag only when it is presented as an IC formed by the purple Ab, for that is when the green epitope becomes accessible to B cells. A similar condition holds for Ag acquisition by purple B cells. The ceiling is now reached when both the B cell lineages have BCRs/Abs with affinities just above $\kappa$. The ceiling is thus still $10^{4}-10^6$ M\textsuperscript{-1}, based on the snapping of a single Fc$\gamma$RIIB-Ab or CR2-Ab complex (Figure 3\textit{b}). That multiple B cell lineages can evolve simultaneously in a GC has been demonstrated.\cite{Tas2016}  

A third scenario results in an effectively bivalent Ag setting when affinity maturation approaches the ceiling (Figure 4\textit{c}). We consider Ags with conformations that preclude the presenting IC from being bound to more than one receptor, either Fc$\gamma$RIIB-Ab or CR2-Ab, on the FDC. The Ag may have multiple, distinct, simultaneously accessible epitopes. The ceiling in this scenario is reached when the affinity of the presenting IC crosses $\kappa$. Because the Ag can be presented by an IC associated with any of the epitopes on the Ag, the ceiling is reached when all B cell lineages involved, each lineage associated with one of the epitopes, evolve to have BCR affinities of $\kappa$. A single BCR is then required to acquire the Ag, rendering the scenario effectively bivalent. The ceiling is again $10^{4}-10^6$ M\textsuperscript{-1}.       

Indeed, with large Ag, a ceiling of $\sim 10^{5}$ M\textsuperscript{-1} has been experimentally observed.\cite{Oda2006} The large Ag may prevent the Ab in the presenting IC from binding additional receptors on the FDC. The flexibility of the Ab hinge region may also constrain the Ab interactions, as demonstrated recently using DNA origami\cite{Shaw2019}, and affect affinity maturation. 

These bivalent, or effectively bivalent, scenarios (Figure 4\textit{a-c}) may explain the low antibody ceilings, smaller than $\sim10^6$ M\textsuperscript{-1}, observed in some situations \textit{in vivo} (Figure 1\textit{b}).

\subsection*{Multivalent antigen}

We now consider the more complex scenario where Ags contain multiple Ab binding sites (or epitopes) and can bind Abs that tether them to the FDC surface via multiple receptors (Figure 4\textit{d-g}). The simplest of these scenarios is when an Ag has three identical Ab binding sites, all simultaneously accessible but sufficiently separated that each is accessible to a distinct BCR/Ab and where each Ab is tethered to the FDC via a single Fc$\gamma$RIIB or CR2 receptor (Figure 4\textit{d}). We again assume a single B cell lineage evolving, so that the Abs in the presenting ICs and the BCR acquiring Ag have the same affinities when the ceiling is reached. The ceiling is now reached when this latter affinity just crosses the overall affinity of the tethers, allowing the BCR to acquire Ag by rupturing the Fc$\gamma$RIIB-Ab and/or CR2-Ab complexes. With three Ab binding sites, the tethers can have a maximum collective affinity of $\kappa^2$, corresponding to two Fc$\gamma$RIIB-Ab and/or CR2-Ab complexes and leaving the third site for BCR binding (Figure 4\textit{d}). The ceiling would thus be $10^{8}-10^{12}$ M\textsuperscript{-1}.

When the three epitopes are distinct, allowing evolution of distinct B cell lineages, the ceiling is again $10^{8}-10^{12}$ M\textsuperscript{-1}, and is achieved when the affinities of BCRs from each of the B cell lineages exceed $\kappa^2$. The ceiling would remain $10^{8}-10^{12}$ M\textsuperscript{-1} in all other configurations where two tethers would have to be broken for Ag acquisition (e.g., see two such configurations in Figures 5\textit{e} and \textit{f}). This affinity ceiling is consistent with the high affinity ceilings observed \textit{in vivo} (Figure 1\textit{b}).

\begin{figure}[!htp]
	\centering
	\includegraphics[width=0.9\textwidth]{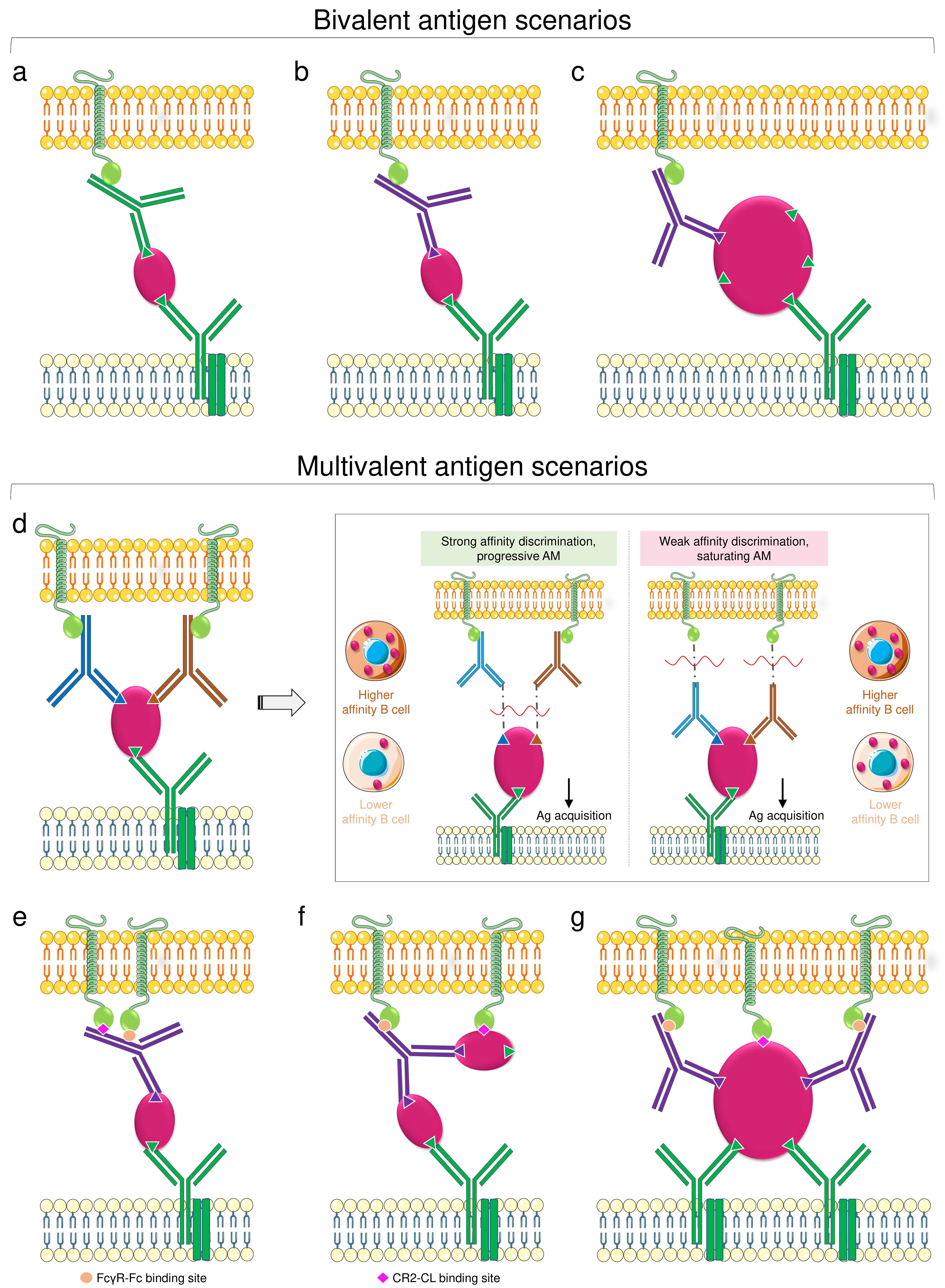}
	\caption{(Continued on the following page.)}
\end{figure}

\addtocounter{figure}{-1}
\begin{figure}[!htp]
	\centering
	\caption{\textbf{Affinity ceiling under different Ag presentation scenarios.} Bivalent Ag with (\textbf{a}) identical, and, (\textbf{b}) distinct epitopes, tethered to the FDC by a single Fc$ \gamma $RIIB/CR2 receptor. (\textbf{c}) An effectively bivalent Ag scenario where a large Ag with many epitopes is bound to the FDC by a single Fc$ \gamma $RIIB/CR2 receptor. Saturating and progressive AM regimes for (\textbf{a-c}) are depicted in Figure 3\textit{b}. (\textbf{d}) Multivalent Ag presented by two Abs tethered to distinct Fc$ \gamma $RIIB/CR2 receptors, along with progressive (left) and saturating (right) affinity maturation illustrated for this configuration. (\textbf{e}, \textbf{f}) Scenarios where the presenting Ab in the IC is linked to the FDC surface by two Fc$ \gamma $RIIB/CR2--Ab tethers. (\textbf{g}) A hypothetical scenario with a highly networked topology of FDC receptors, Abs, Ag and BCRs.}
\end{figure}  

In general, we can imagine Ag being held by `$\phi$' tethers and tugged by `$\beta$' BCRs (Figure 4\textit{g}). The ceiling is reached when with collective affinity (or avidity) of the tethers equals the total affinity of the bound BCRs. If the affinity of a BCR is $ \rho\left(=K_a\right) $, the balance implies $\kappa^{\phi}=\rho^{\beta}$, so that the ceiling becomes $\left({\kappa}\right)^{\phi/\beta}$, with $\kappa=10^{4}-10^6$ M\textsuperscript{-1}. In Figure 4\textit{g}, $\phi=3$ and $\beta=2$, thus leading to a ceiling of $10^{6}-10^9$ M\textsuperscript{-1}. It follows that the ceiling rises as $\phi$ increases and $\beta$ decreases. Thus, the ceiling is the largest when $\beta=1$ and $\phi$ corresponds to the maximum tethers that can be accommodated. We speculate here, assuming $\beta=1$, that either increasing $\phi$ beyond 2 is sterically hindered or makes the force required for Ag extraction so large that B cells switch to an affinity-independent enzymatic Ag extraction pathway\cite{Spillane2017,Yuseff2011} restricting the ceiling to $10^{8}-10^{12}$ M\textsuperscript{-1}. 

\section*{Discussion and outlook}        

The weakest link hypothesis is similar in spirit to the explanation of the affinity ceiling offered first by Foote and Eisen\cite{Foote1995}. The ingenuity of their explanation was in its recognition that the ceiling would arise from processes extrinsic to the specific Ag-Ab interactions. With soluble Ag, these processes were diffusion and internalization of Ag. With surface tethered Ag, which we consider, the ceiling comes from the Fc$\gamma$RIIB-Ab or CR2-Ab interactions. \textit{In vivo}, because Ag acquisition in GCs is predominantly from the FDC surface, the weakest link hypothesis is expected to prevail. In addition, by considering different possible configurations of Ag presentation to GC B cells, we show how the wide range of maximum affinities observed \textit{in vivo}, from $\sim 10^{5}-10^{12}$ M\textsuperscript{-1}, could be realized.

Evidence in support of the weakest link hypothesis comes from \textit{in vitro} experiments that modify the weakest link. In elegant experiments, Spillane and Tolar{\cite{Spillane2017}}, attached Ag covalently via DNA tethers and streptavidin tethers to two kinds of surfaces: 1) stiff and immobile (glass modified with polyethylene glycol; termed as `PEG') and 2) flexible and partly mobile (plasma membrane sheets; termed as `PMS'). B cells interacted with this Ag, thus, through a chain of three complexes: Ag-DNA, DNA-streptavidin, and streptavidin-surface. Ag extraction by B cells was compared between the stiff PEG and the flexible PMS surfaces. When Ag was presented on the stiff PEG surface, B cells preferentially ruptured the weaker DNA-streptavidin complex and acquired Ag. However, with the flexible PMS, which appeared to be the weakest link, the entire chain of complexes along with an annular surface patch was internalized and resulted in similar amounts of extracted Ag bound to either weak or strong DNA tethers, marking a loss of affinity discrimination and a corresponding lowering of the affinity ceiling to the strength of the membrane patch. 

More direct tests of the hypothesis could include mutating the Fc region of Abs so that the affinity of the weakest link in the chain could be modified. Mutagenesis of Fc$\gamma$RIIB shows that such modifications are possible{\cite{Shields2001}}. Similarly, an Ab with mutations in the Fc region showed an $\sim$200-fold higher affinity for the Fc$\gamma$RIIB receptor.{\cite{Mimoto2013}} Several studies have explored CR2 and C3d mutants which either stabilize or destabilize the CR2-C3d interaction.{\cite{vandenElsen2011,Hannan2005}} Corresponding changes in the ceiling would validate the weakest link hypothesis. It would be interesting to examine whether the humoral responses of individuals with clinically relevant, naturally occurring polymorphisms of the Fc$\gamma${\cite{Mellor2013,Sakai2017}} and CR2 {\cite{Wu2007}} exhibit different affinity ceilings. 

While we have resorted to estimates of $ \Delta G $'s of the links under various Ag presentation scenarios to quantify the ceiling, we recognize that these estimates are subject to uncertainties arising from the behavior of the links under tensile forces. Mechanical forces can change the on and off rates in complex ways. In `slip bonds', tensile forces hasten dissociation, whereas in `catch bonds', they may delay dissociation; some bonds can exhibit both behaviours depending on circumstances{\cite{Rakshit2012}}. \textit{In vitro} experiments suggest that the BCR-Ag interaction is a complex slip bond, where the dissociation rate increases steeply for small forces and less steeply for large forces.{\cite{Natkanski2013,Tolar2014_book}} The nature of the other links in the chain is less well described, precluding precise estimation of the ceiling. Nonetheless, Batista and Neuberger have argued that despite the mechanical forces involved, Ag acquisition by B cells may still depend on the `intrinsic' quality of the BCR-Ag interaction\cite{Batista2000}, justifying the use of $ \Delta G $'s to obtain at least an approximate quantification of the ceiling. 

Ab-mediated immune responses depend strongly on the affinity of the Abs for their target Ag.{\cite{Maynard2002,Carter2006,Brandenberg2017,Lu2018}} It is important, therefore, to design vaccination and other intervention protocols such that the highest affinity Abs can be elicited. The affinity ceiling presents a natural limit to such elicitation. Our study presents insights into the ceiling that may suggest ways of overcoming the limit. One strategy could involve passive immunization with engineered Abs. Passive immunization with broadly neutralizing Abs of HIV-1 have been found to improve the endogenous humoral response to HIV-1.\cite{Schoofs2016} A plausible mechanism underlying this improvement is that the passively administered Abs enter GCs via the Ab feedback mechanism and improve selection stringency.\cite{Zhang2013, Garg2019} Modifying the Fc regions of the passive Abs to increase their Fc$\gamma$RIIB binding affinity, possibly also using tight binding IgG subtypes\cite{Bruhns2009}, would strengthen the weakest link, potentially raising the affinity ceiling and further improving the humoral response. Such interventions may take us a step closer to more potent vaccination strategies, including those leading to a functional cure of HIV-1 infection.{\cite{Freund2017}}\newline

\section*{Acknowledgements}

We thank the Wellcome Trust/DBT India Alliance Senior Fellowship IA/S/14/1/501307 (NMD) for funding. We acknowledge Peter Hraber for inputs on the affinities of HIV-1 broadly neutralizing antibodies.\newline

\section*{Author Contributions}

R.D. conceptualized the study. R.D., R.A. and N.M.D. analysed results and co-wrote the manuscript.\newline

\section*{Competing interests}

The authors declare no competing interests.\newline

\bibliography{Refs}
\bibliographystyle{naturemag}

\end{document}